\documentclass[twocolumn,floatfix, showpacs]{revtex4}

\usepackage[final]{epsfig}
\usepackage{amsmath}

\begin{document}
 
\title{The space-time structure of the energy deposition into the bulk medium due to jet quenching}
 
\author{Thorsten Renk}
\email{thorsten.i.renk@jyu.fi}
\affiliation{Department of Physics, P.O. Box 35, FI-40014 University of Jyv\"askyl\"a, Finland}
\affiliation{Helsinki Institute of Physics, P.O. Box 64, FI-00014 University of Helsinki, Finland}

\pacs{25.75.-q,25.75.Gz}

\begin{abstract}
While jet quenching in ultrarelativistic heavy-ion collisions is firmly established as a phenomenon resulting from the interplay between hard perturbative and soft fluid-dynamical Quantum Chromodynamics (or equivalently the interaction between hard probes and bulk QCD matter), less is known with certainty about the reaction of bulk matter to the passage of a jet. On general grounds, a jet interacting while passing through a medium represents a source of energy and momentum for the bulk matter fluid. If the precise form of such a source term is known, the reaction of the medium can be computed using fluid dynamics. Recent advances in the understanding of hard probes due to the wealth of data from RHIC and LHC allow to constrain the source term better by determining the energy flow away from hard modes. The aim of this work is to discuss what can be learned from such constraints in the context of the in-medium shower evolution code YaJEM-DE and to illustrate the role of fluctuations in the energy deposition.
\end{abstract}
 
\maketitle

\section{Introduction}

Several years ago, the PHENIX collaboration at RHIC observed the splitting of the away side correlation peak in triggered dihadron correlation from a broad Gaussian to a double-hump structure when going from peripheral to central Au-Au collisions \cite{Double-Hump}. As a mechanism responsible for this structure, Mach cones excited in the medium by the deposition of energy due to the energy loss of a hard parton in a partonic back-to-back event were quickly suggested \cite{Stocker,Solana,Wake} and more detailed studies of the phenomenology showed that this explanation was at least consistent with the observed phenomenon when certain assumptions about the coupling of jet to the medium are made \cite{Mach1,Mach2,Mach3}. However, it was later shown that such double-hump structures in low $P_T$ triggered correlation are  generated in a much more natural way by event by event (EbyE) fluctuations in the hydrodynamical initial state, in particular by fluctuation-driven triangular structures \cite{Alver,Hannah}. In addition, jet-h correlations as measured by the STAR collaboration \cite{STAR-jet-h} where a substantial amount of energy loss into the medium (and hence available for shockwave excitation) was expected did not lead to any obserable double-hump structure. This significantly diminished the interest in the energy deposition into the medium caused by jet quenching and the strategy of determining the medium speed of sound by measuring a Mach cone opening angle.

Nevertheless, there is growing evidence across several models that energy flow from perturbative jet-like to non-perturbative medium-like degrees of freedom is needed to account for both the dijet asymmetry observed in Pb-Pb collisions at the LHC \cite{Collimation,Dijets-Qin,Dijets-Renk} and for the associate momentum dependent of hard dihadron correlations \cite{IAA_elastic}. Thus, while energy-loss driven hydrodynamic excitations were not a suitable explanation for the observed correlation phenomena, it appears that the concept of energy deposition by a hard parton shower into the medium as such and the idea of the excitation of hydrodynamical modes are sound. Modelling the hydrodynamical response to such an energy deposition might be crucial to understanding the correct background to a medium-modified jet, since this mechanism would generate a medium background correlated with a jet.

It is however known that the expected hydrodynamical response to energy deposition depends strongly on what is assumed for the space-time structure of the source term. Various assumptions have been made in the literature, for instance in \cite{Mach1} the space-time dependence of energy deposition into the medium was computed from leading parton energy loss in the Armesto-Salgado-Wiedemann (ASW) formalism \cite{QuenchingWeights} and found to be a function peaked around 3-4 fm whereas in \cite{Betz} a Bethe-Block source term with a Bragg peak at the end of the energy deposition phase has been assumed. According to a case study with different source terms in linearized hydrodynamics \cite{Neufeld}, the observable medium response is strongly dependent on the assumed space-time structure of the source and a double-hump structure is only pronounced when the fluid viscosity is low and the energy deposition is peaked towards the end.

The aim of this paper is to estimate the space-time structure of energy deposition into the medium from the high $P_T$ side, i.e. to compute average and EbyE fluctuating energy deposition using the multiple-observable constrained \cite{Systematics} in-medium shower evolution code YaJEM-DE \cite{YaJEM1,YaJEM2,YaJEM-DE}

\section{Jets in medium}

For a jet as created in e.g. an $e^+e^-$ collision in which there is no background medium, a (calorimetric) measurement of the energy flow inside a jet cone is approximately equivalent to the sum of  shower parton energies or hadron energies detected inside the cone. This is a consequence of energy-momentum conservation --- the evolution of the initial highly virtual partons into a Quantum Chromodynamics (QCD) shower takes place in vacuum, and hence energy and momentum must be conserved inside the evolving partonic (hadronic) system.

This is manifestly not so in the presence of a background medium. Here, energy and momentum are conserved for the whole heavy ion collision event, but once one subdivides the event into a non-perturbative evolution of the thermalized bulk medium and a perturbative QCD (pQCD) evolution of partons generated by hard processes inside the medium, energy and momentum is not conserved separately inside the perturbative and the non-perturbative sector. One consequence of this is that the energy-momentum flow of the jet inside a cone is in general no longer recoverable by the sum of energies of selected hadrons, since the energy of any particular hadron may come partially from the jet and partially from the medium temperature. In particular, energy dissipated into bulk, hydrodynamic excitations is effectively shared across many ($O(100-1000)$) hadrons.

This is particularly relevant when interpreting theoretical modelling of heavy-ion collisions where the calculational techniques for bulk and hard processes are vastly different: While the bulk is usually treated by solving the equations of ideal or viscous relativistic fluid dynamics (see e.g. \cite{hydro1,hydro2,hydro3,hydro4,hydro5}), the evolution of in-medium parton showers into jets is typically treated by pQCD-based Monte-Carlo (MC) modelling \cite{YaJEM1,YaJEM2,YaJEM-DE,JEWEL,Q-PYTHIA,MARTINI}. In the absence of a consistent framework to treat both bulk and hard processes on the same footing, this implies that a medium-modified shower model may violate energy-momentum conservation inside the perturbative sector by coupling to a bulk medium which is not explicitly modelled (is is also true that fluid dynamical calculations might in principle be allowed to violate energy-momentum conservation by coupling to a perturbative sector which is not explicitly modelled, however the typical energy of even an energetic jet is small when compared to the whole energy stored in the bulk medium). This is not true in frameworks where the medium modification of a shower is assumed to change only splitting probabilities (such as for instance Q-PYTHIA \cite{Q-PYTHIA}) in which case energy-momentum conservation is exact in the perturbative sector alone. This, however, can not be assumed  \emph{a priori} and is not justified by any microscopic physics picture.

Physical intuition argues that in general the energy flow should be from hard, perturbative modes into soft, non-perturbative modes since the hard modes are expected to thermalize and become part of the bulk in the limit of sufficiently long times spent in the medium. However, in the short term, the energy flow may be reverse, for instance a hard parton may scatter with the bulk medium and produce a recoiling parton which is sufficiently energetic to be considered hard by itself \cite{JEWEL,EMC}. If the recoiling parton is then formally counted into the perturbative sector and evolved with the pQCD shower, this corresponds to an energy transfer from the bulk medium to the perturbative shower (note that in this example it is still true that momenta in general soften --- however the average of a hard and a soft momentum may lead to a situation in which both final state momenta are counted as hard, despite being softer than the initial state hard momentum). 

It is therefore clear that making the concept of non-perturbative bulk vs. perturbative hard sector quantitative requires a scale separation. Considering a gluon with energy $E$ of order of the medium temperature $T$ a perturbative object is meaningless, since it is indistinguishable from bulk gluons, so clearly $E \sim \text{few } T$ must be a minimal criterion for a perturbative degree of freedom. On the other hand, the separation scale does not seem to be so much a theoretical concept but rather a physical transition scale at which the behaviour of the system undergoes pronounced changes: While pQCD shower dynamics is dominated by singularities in the gluon emission kernels leading to the dominance of soft and collinear (i.e. forward) scattering and particle production, the validity of hydrodynamics implies fairly complete isotropization and hence the complete loss of any pQCD-specific dynamics. Gaining a precise understanding of where the separation scale is and how physics changes in its vicinity is one of the most exciting current challenges in the field of ultrarelativistic heavy-ion collisions. 

In the following, we start tackling this question from the hard sector by posing the question how the energy-momentum balance in the perturbative sector is realized in a model that is constrained by data and what follows for the energy-momentum balance in the medium. We start with the simplest case of a drag force acting on partons in a constant medium and develop a picture of the full problem from there.

\section{Constant medium}

\subsection{General considerations}

We consider the bulk medium for the purpose of interaction with hard, perturbative partons to be characterized completely by transport coefficients. Here, $\hat{e}$ represents the mean momentum loss per unit pathlength $x$ along the direction of the parton $dp_z/dx$ (which for massless partons is to good approximation equal to the energy loss) and $\hat{q}$ represents the virtuality transfer per unit pathlength $dQ^2/dx$. We assume in the following that energy lost via $\hat{e}$ is completely dissipated into the medium.

The simplest case is that of a massless on-shell parton with initial energy $E_0$ traversing a medium of length $L$ characterized by a constant value of $\hat{e}$ and $\hat{q}=0$. In the case  $\hat{e} L \ll E_0$ the parton has effectively infinite energy and the mean energy loss into the medium (i.e. the energy deposition) per unit length is given by $\langle dE/dx \rangle = \langle dp_z/dx \rangle = \hat{e}$. However, the actual energy loss may fluctuate around the mean value, and thus there is a finite and pathlength-dependent survival probability $P_S(x)$ for a parton to be still present beyond $\hat{e} L = E_0$, i.e. the point where partons on average have lost all their energy. For $L > E_0/\hat{e}$, the energy deposition is thus rather determined by the expected number of surviving partons and thus $dE/dx \sim P_S(x) \hat{e}_1$ where $\hat{e}_1$ in general not equal $\hat{e}$ since there may be a bias for surviving partons. This difference is dependent on the precise nature of the fluctuations --- if the energy deposition is constant along the path but with a different value per parton, only partons with small energy loss survive. If energy loss is fluctuating along the path randomly, no such bias exists.

However, in a real  hard QCD process, partons in the out state are highly virtual and this leads to the evolution of a partonic shower from a parent parton into multiple daughter partons. Thus, if a virtual initial parton with energy $E_0$ is placed into a constant medium, the energy-depositing sources quickly multiply due to the development of a parton shower ('vacuum branching') and the energy deposition for short pathlengths becomes parametrically $\langle dE/dx \rangle \sim N_{part}(x) \hat{e}$ where $N_{part}(x)$ is the mean number of partons in the shower which have decohered sufficiently from the parent to scatter independently after the shower has evolved for a length $x$. The interplay between parton (de)-coherence and the ability of the medium to resolve individual partons is important for a correct treatment of the problem (see e.g. \cite{Coherence} for a discussion).

Even if the initial parton is on-shell, a finite value of $\hat{q}$ in a medium is capable of inducing additional gluon radiation from a parent parton as well as from daughter partons \cite{radiative1,radiative2,radiative3,radiative4,radiative5,radiative6}. Thus, in a medium characterized by finite $\hat{e}$ and $\hat{q}$ the medium itself quickly increases $N_{part}(x)$ with $x$ even for an on-shell initial parton and the total energy deposition grows with $x$ --- this scenario is known as the 'Crescendo' \cite{Crescendo}.

A Crescendo scenario can however only persist as long as $\hat{e} x < E_i$ for every shower parton energy $E_i$. If this condition is not met for a parton, the parton is likely to be absorbed by the medium and is no longer available as a source for energy deposition. Such finite energy corrections counteract the Crescendo effect, and since gluon radiation as computed in pQCD tends to be parametrically soft, finite energy corrections apply early for radiated daughter partons. In practice, it was found that the energy deposition for a shower with vacuum branching and finite energy corrections is decreasing \cite{Neufeld} with $x$ over most of the range, as initial vacuum branchings quickly increase $N_{part}$ to a peak value, and then finite energy correction dominate and the total energy deposition goes down as less and less surviving partons are found at large $x$.

Dependent on how precisely the kinematics is realized, there is an additional complication: The action of $\hat{q}$ widens the shower in transverse space by overcoming kinematical constraints which are present without a medium. This widening in general takes energy which may come both from the leading parton and from the medium. Such energy flow is usually hidden in analytical models where the hard parent parton is assumed to have infinite energy and the medium is assumed to consist of static scattering centers, but in principle the action of $\hat{q}$ also alters the energy balance between jet and medium and may increase the energy in the hard modes. %We will discuss this contribution later.

\subsection{Results}

In order to illustrate these ideas, we extract the energy deposition into a constant medium from the in-medium shower evolution code YaJEM \cite{YaJEM1,YaJEM2,YaJEM-DE}, in a first run with $\hat{q}$ set to zero and $\hat{e}$ the only relevant coefficient (this is referred to as 'YaJEM-E'). YaJEM is based on the PHSHOW code \cite{PYSHOW} (to which it reduces in the absence of a medium) and is primarily designed to simulate the evolving shower, hence it has limited option to simulate the detailed fluctuation pattern of energy deposition for a single parton: In YaJEM-E, $\Delta E = \hat{e} L_i$ (where $L_i$ is the length traversed by the virtual shower parton $i$) is allowed to fluctuate around the mean value, but the energy deposition is taken to be constant along the path segment $L_i$ (whereas in principle at this point phenomena like a Bragg peak leading to a change in $\hat{e}$ along the path segment might be relevant). The primary justification for using this fairly simple model of energy deposition from a single parton is that other EbyE fluctuation-generating effects such as the fluctuation of $N_{part}$ or of the parton formation times \cite{YaJEM2} which are included in the model already are quite substantial (see below), and thus the faithful simulation of fluctuations in the path dependence of the energy deposition of a single parton may not be that essential.

\begin{figure}[htb]
\begin{center}
\epsfig{file=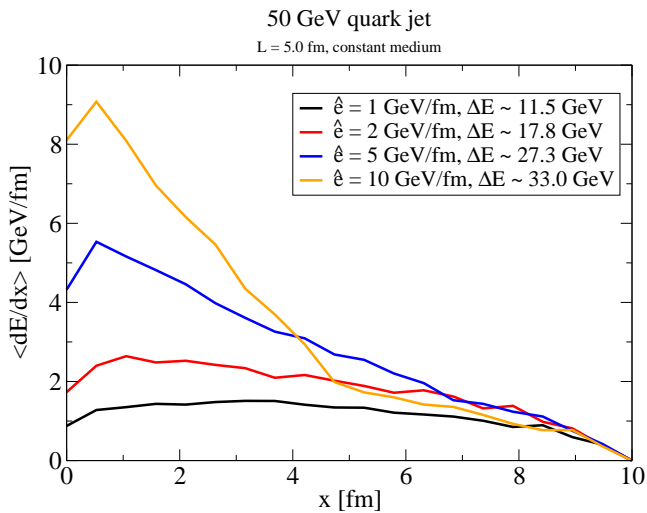, width=8.5cm}
\end{center}
\caption{\label{F-constant}(Color online) Energy deposition of a parton shower initiated by a 50 GeV quark into a medium characterized by a constant transport coefficient $\hat{e}$ for different values of $\hat{e}$.}
\end{figure}

Results for the energy deposition of a 50 GeV quark jet under the assumption that $\hat{q} = 0$ in a constant medium with 10 fm length are shown in Fig.~\ref{F-constant}. These reproduce the 'decreasing' source term computed in \cite{Neufeld} in the case where most of the jet energy is lost into the medium. Without substantial finite energy corrections, for $\hat{e} = 1$ GeV/fm, the energy deposition is fairly constant across the whole range. The total amount of lost energy $\Delta E$ is larger than the value $\hat{e} L = 10$ GeV one would expect for a single parton, i.e. on average there is more than one energy-depositing source. The role of the vacuum shower evolution generating additional sources can also be seen in the small initial rise of $\langle dE/dx \rangle$.

For larger values of $\hat{e}$, finite energy corrections become important and the peak of the energy deposition moves more and more towards early times. The presence of strong fluctuations in $\Delta E_i$ ensure that the energy deposition is non-zero even at large $x$ where the average energy deposition exceeds the shower energy. In combination with finite energy corrections, fluctuations ensure that the total energy deposition can never grow above the shower energy $E_0$ no matter the value of $\hat{e}$.

Thus, the main physics mechanisms discussed above, such as the effect of the survival probability or the average parton number can be identified in the model results. We now turn to the more realistic case of an evolving medium with a finite $\hat{q}$, but still neglect the explicit effect of $\hat{q}$ on the energy balance.

\section{Hydrodynamical medium}

\subsection{Basic structure of the results}

In an expanding medium as generated in a heavy-ion collision, $\hat{q}$ and $\hat{e}$ become functions of the spacetime position of the hard parton. These functions depend on the distribution and evolution of bulk matter and on the initial hard vertex position and transverse direction of the back-to-back parton event. In general, there is an infinity of configurations. In \cite{YaJEM2} it was shown that for a majority of these paths transport coefficients can be parametrized with a simple power law, and that moreover the medium modification of the shower largely depends on the line-integrated virtuality $\Delta Q^2_{tot} = \int d x \hat{q}(x)$ along the path of the shower initiator.

In the following, we compare two characteristic situations, a path from the center of the medium to the surface and a short path from a vertex close to the surface outward for various values of $\Delta Q^2_{tot}$. All results are now obtained using the scenario YaJEM-DE which results in the best agreement with all available high $P_T$ data. This scenario uses a balance of about 10\% of elastic energy transfer to the medium and 90\% induced perturbative ratiation as constrained by a large body of observables \cite{Systematics}. As a rough guide, typical $\Delta Q^2_{tot}$ relevant for RHIC in-medium shower calculations range from 4 to 8 GeV$^2$ and about twice this number for LHC.

\begin{figure}[htb]
\begin{center}
\epsfig{file=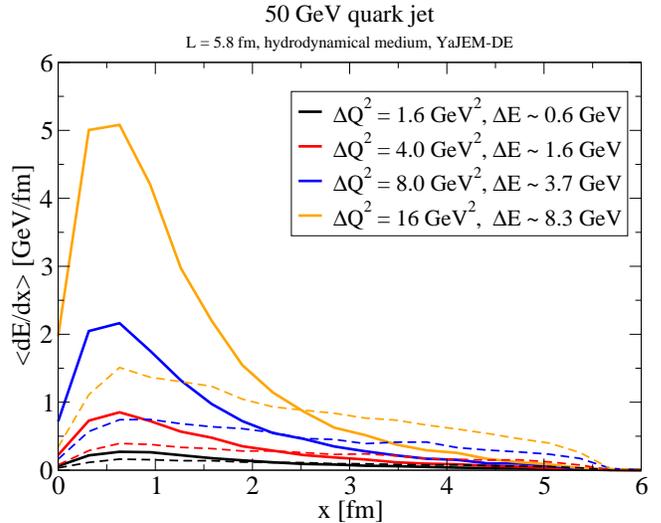, width=8.5cm}
\end{center}
\caption{\label{F-hydro_long}(Color online) Energy deposition of a parton shower initiated by a 50 GeV quark placed into the center of an evolving medium for different values of the line-integrated virtuality transfer $\Delta Q^2$ with the relative strength of $\hat{q}$ and $\hat{e}$ determined by data. }
\end{figure}

Fig.~\ref{F-hydro_long} shows results for a path approximately from the medium center. In comparison to the case of a constant medium, the energy deposition is much more peaked towards early deposition at small lengths $x$. This can naturally be understood by taking into account that the medium density (and hence its transport coefficients) drop due to the expansion of the medium and due to the fact that the parton travelling outward reaches the dilute surface of the medium, hence late-time medium effects are always suppressed in an expanding medium.

The effect of the Crescendo can be seen by comparing solid with dashed lines where for the dashed lines $\hat{q}=0$ has been  assumed. The additional fast multiplication of partons due to medium-induced radiation at early times is clearly seen as a rapid rise at small $x$, especially for large $\Delta Q^2_{tot}$ (coresponding to large $\hat{q}$). However, finite energy corrections in combination with the decreasing medium density quickly reverse the initial steep rise of the mean energy deposition.

\begin{figure}[htb]
\begin{center}
\epsfig{file=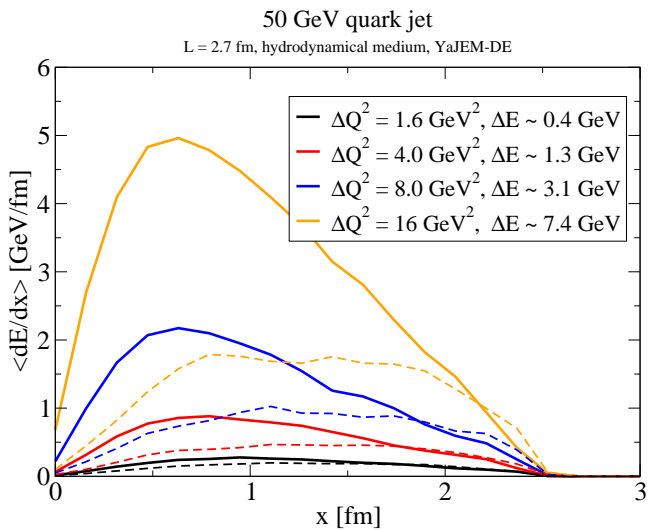, width=8.5cm}
\end{center}
\caption{\label{F-hydro_short}(Color online) Energy deposition of a parton shower initiated by a 50 GeV quark placed at the periphery of an evolving medium for different values of the line-integrated virtuality transfer $\Delta Q^2$ with the relative strength of $\hat{q}$ and $\hat{e}$ determined by data.  }
\end{figure}

Fig.~\ref{F-hydro_short} shows the results for a short pathlength of 2.7 fm, i.e. a shower which reaches the medium surface quickly (note that since $\Delta Q^2_{tot} = \int dx \hat{q}(x)$ the typical values of $\hat{q}(x)$ are higher for a short pathlength if the same $\Delta Q^2$ is considered --- this is important in comparing Figs.~\ref{F-hydro_long} and \ref{F-hydro_short} but trivially addressed in any calculation in which the actual averaging over a hydrodynamical medium is done). 

Qualitatively the scenario is unchanged --- the Crescendo-effect leads to a rapid growth of energy deposition sources which reaches a maximum before being turned over by finite energy corrections and the medium density dilution. Quantitatively, to 0th order the total energy deposition scales reasonably with $\Delta Q^2_{tot}$ independent of the in-medium pathlength, however the pathlength affects the energy deposition on the 30\% level.

It is worth noting from the numbers for the integrated energy loss that at high $P_T$ RHIC and especially LHC kinematical conditions showers are typically not completely absorbed by the medium, but that the larger fraction of the shower energy remains in perturbative modes.

\subsection{The situation at LHC kinematics}

In the kinematic reach currently probed by the LHC  observables (see e.g. \cite{ATLAS,CMS,CMS-Edep}) hard partonic back-to-back events are characterized by a notable fraction of gluon jets. Gluons as shower-initiators show two main differences to quarks: The coupling to the medium is increased by a color factor $C_F = 9/4$ and due to the splitting $g \rightarrow q\overline{q}$ which prefers equal momenta for the quarks, gluon jets have a softer fragmentation and a somewhat broader shape even in vacuum.

\begin{figure}[htb]
\begin{center}
\epsfig{file=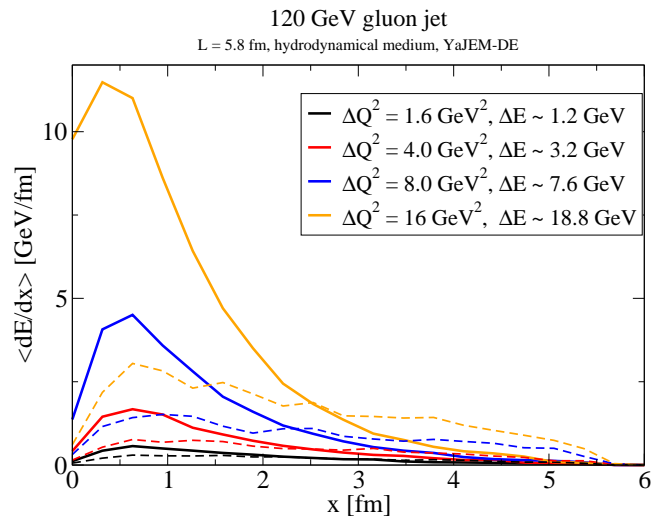, width=8.5cm}
\end{center}
\caption{\label{F-LHC}(Color online)  Energy deposition of a parton shower initiated by a 120 GeV gluon placed into the center of an evolving medium for different values of the line-integrated virtuality transfer $\Delta Q^2$ with the relative strength of $\hat{q}$ and $\hat{e}$ determined by data. }
\end{figure}

The results for a 120 GeV gluon as a shower initiator are shown in Fig.~\ref{F-LHC}, again for a longer path of 5.8 fm corresponding to partons produced roughly in the medium center. Qualitatively, the results are very similar to the case of a 50 GeV quark for the same pathlength. Quantitatively, the scaling of the total energy deposition is roughly consistent with the different color factor. 

There is however a problem with this interpretation: The similarity of the functional shape of the energy deposition suggests that the information of the shower-initiationg parton type is quickly made obsolete by abundant medium-induced radiation. In this case the Casimir factor should not be reflected in the results. 

\begin{figure}[htb]
\begin{center}
\epsfig{file=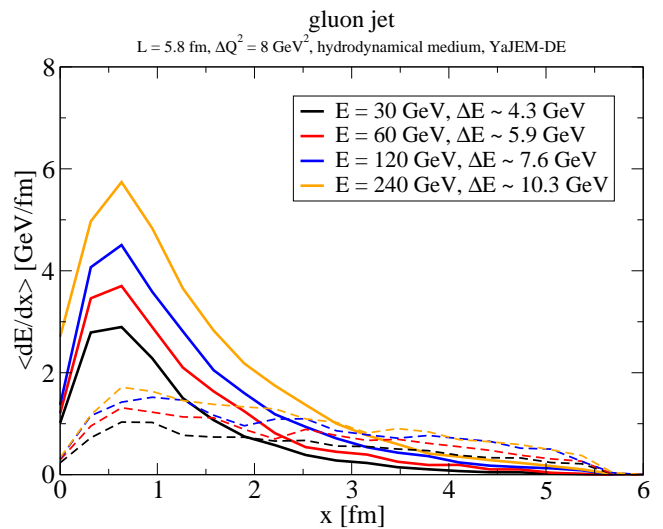, width=8.5cm}
\end{center}
\caption{\label{F-LHC-E0}(Color online)  Energy deposition of a parton shower initiated by a gluon placed into the center of an evolving medium for different values of the initial gluon energy $E_0$ with the relative strength of $\hat{q}$ and $\hat{e}$ determined by data. }
\end{figure}

In order to clarify the situation, in Fig.~\ref{F-LHC-E0} the dependence of the energy deposition on the initial parton energy $E_0$ is shown. This mainly affects how soon finite energy correction become relevant. The dependence of the total mean energy deposition on initial parton energy can be well fit by $\Delta E \sim \frac{E_0}{1 \text{GeV}}^{0.37}$. This suggests that at good part of the normalization difference between Figs.\ref{F-hydro_long} and \ref{F-LHC} is due to the difference in $E_0$, which is confirmed by an explicit calculation.

\section{Event-by-event fluctuations}

\subsection{Fluctuation sources} 

There are multiple sources for event-by-event fluctuations around the mean energy deposition of a shower given an in-medium path. They can broadly be grouped into the following categories:

\begin{itemize}
\item fluctuations of the energy deposition of single partons along their path 
\item fluctuations of $N_{part}(z)$ in the shower evolution
\item fluctuations in the background medium density, translating into fluctuations of the transport coefficients
\end{itemize}

The approximate scaling of medium effects with $\Delta Q^2_{tot}$ identified in \cite{YaJEM2} and explicit calculations in \cite{Fluct} suggest that fluctuations in the medium density are a subleading effect. On the other hand, the relative strength of the Crescendo effect observed in Figs.~\ref{F-hydro_long},\ref{F-hydro_short} and \ref{F-LHC} above the baseline calculations that contains already fluctuations in the energy deposition of single partons suggests that particle numbers are large and the dominant effect are fluctuations in $N_{part}(z)$ which are captured  by YaJEM.

\begin{figure}[htb]
\begin{center}
\epsfig{file=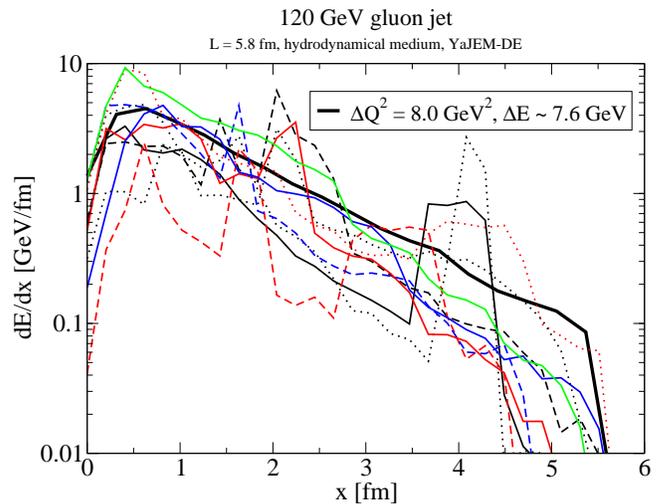, width=8.5cm}
\end{center}
\caption{\label{F-fluct}(Color online)  Energy deposition of a parton shower initiated by an 120 GeV gluon placed into the center of an evolving medium, shown both as mean value and for 10 individual shower events. The relative strength of $\hat{q}$ and $\hat{e}$ is determined by data. }
\end{figure}

\subsection{Results}

In Fig.~\ref{F-fluct}, the mean energy deposition of a 120 GeV gluon is shown along with the energy deposition in 10 individual events. The fluctuations are fairly strong, up to a factor three different from the average, and thre relative strength of fluctuations persists during the whole evolution. Upward spikes in the energy deposition can clearly be seen and identified as the emission of a daughther parton to the point that it is resolved by the medium where the length in $x$ of the upward spike correlates with the energy of the daughter parton and the (fluctuating) $\hat{e}$ governing its energy loss --- as soon as a daughter parton energy is depleted, the total energy deposition decreases again.

The strong fluctuations seen in this result argue that in order to have a realistic picture of energy deposition into the medium, the average energy deposition is not sufficient and EbyE fluctuations need to be taken into account.

\section{Scale separation and energy balance}

Let us now return to the effect of $\hat{q}$ on the energy balance. In YaJEM, a shower gains the energy for transverse broadening largely from the medium. The microscopical interpretation of this is that medium partons  are being 'swept away' by the shower and hence become correlated by the jet, thus if their energy is formally counted as part of the jet, the in-medium jet energy keeps growing \cite{YaJEM2}.

As mentioned before, this is not a reasonable physical interpretation, because there is no physical distinction between soft medium and soft jet gluons, and hence soft gluons can not be counted as part of a perturbative jet inside a medium. For a proper interpretation, we need to introduce a separation scale between hard perturbative and soft fluid-like physics below which partons are counted as part of the medium. Note that there's an implicit assumption involved that the medium is strongly interacting and manifestly not perturbative below the separation scale --- with just a separation scale selected, even a vacuum shower would lead to a positive energy deposition for the simple reason that some radiated gluons would fall below the separation scale, however no such reasoning is justified since the emission of soft gluons appears to remain sufficiently perturbative in vacuum. The assumption is hence that soft gluons would not only fall below the separation scale but  also be subject to the physics conditions below the scale, i.e. they would be isotropized just as the rest of the bulk medium.  

\begin{figure}[htb]
\begin{center}
\epsfig{file=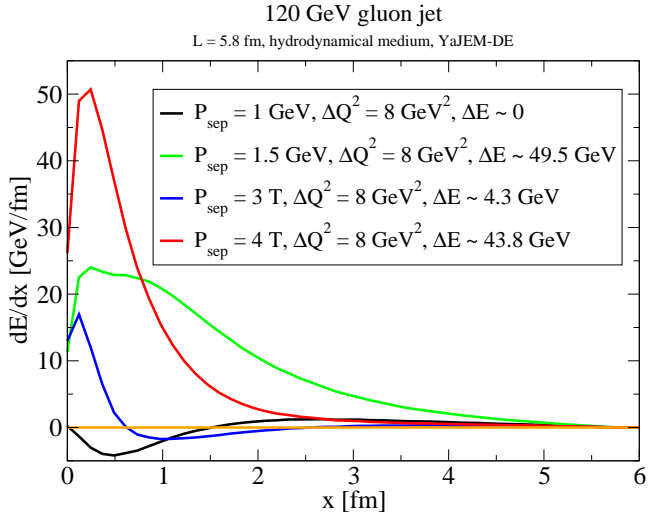, width=8.5cm}
\end{center}
\caption{\label{F-sep}(Color online)  Mean energy deposition of a parton shower initiated by an 120 GeV gluon placed into the center of an evolving medium with  different scale separation schemes taken into account (see text). }
\end{figure}

\emph{A priori} the choice of the separation scale is not unique. We might think for instance of a fixed momentum scale or a multiple of the system temperature $T$. In Fig.~\ref{F-sep}, the resulting energy deposition into the non-perturbative system is shown for different choices of the separation scale.

For a constant $P_{sep}$ of 1 GeV, the energy deposition turns initially negative (indicating that parts of the medium are swept into a perturbative region) before turning over to a positive value at late times, by chance cancelling out to a vanishing net energy deposition. For a scale choice of $P_{sep}=1.5$ GeV on the other hand, the energy deposition is positive throughout and assumes a large value.

If the separation scale is chosen as a multiple of the temperature, energy deposition is enhanced at early times (where $T$ is large) and suppressed at late times as compared with a constant $P_{sep}$. An intermediate negative region is found for $P_{sep} = 3T$ whereas for $P_{sep} = 4T$ the energy deposition remains positive in the entire region.

It is very clear from these results that the separation scale choice has significant influence on the interpretation of the results. However, unlike many other scale choices (for instance a renormalization scale, or the scale governing the transition from parton shower dynamics to hadronization in computations of the fragmentation function), the physics in this case is not expected to be approximately independent of the scale choice in some transition region. Rather, the system created in heavy-ion collisions appears to be characterized by a sharp transition over a narrow momentum range from fluid-like to jet-like behaviour, with interesting phenomena such as recombination \cite{Reco,Coalescence} characterizing the transition region. Thus, the nature of $T_{sep}$ contains information of how the system makes the transition from a weakly coupled to a strongly coupled state and must thus be obtained by comparison of theoretical scenarios with measurements. A clear experimental signature for the change from jet-like to medium-like degrees of freedom carrying the shower initiating parton energy is thus the change from jet-like to medium-like hadrochemistry, as seen e.g. in the thermal enhancement of multi-strange hadrons which has already been observed and discussed long ago at SPS energies \cite{Hadrochemistry}.

\section{Discussion}

The results obtained in this work illustrate that the energy deposition by a hard evolving jet into a bulk fluid-like medium is by no means a simple question. The Crescendo effect requires to discuss the problem in terms of a parton shower rather than leading parton physics only, this in turn implies that fluctuations around the average driven by the spacetime structure of the shower evolution are substantial. Approximating the problem by a mean energy deposition may therefore be insufficient. A substantial additional complication is that in addition to the explicit energy deposition into the medium via $\hat{e}$, there is also an implicit deposition having to do with maintaining a separation scale between hard jet constituents and soft medium constituents. The number for the formally counted energy deposition into the soft sector can be shown to be extremely sensitive to the precise scale separation scheme. This means that further input to determine around what scale the behaviour of matter changes from the characteristic pQCD forward scattering to fluid-like isotropic scattering in strong coupling is needed. 

Nevertheless, a common result of almost all scenarios considered here is that the energy deposition is characterized by a rapid initial rise, followed by a quick decay to small values at large time. This means that the main perturbation due to jets occurs when the medium itself is characterized by significant event-by-event fluctuations from the initial state \cite{Alver,Hannah}, and that viscosity will in all likelihood wash out any specific Mach-cone like signal in the subsequent evolution \cite{Neufeld}. The most likely observable signature of a medium reaction to energy deposited by a jet is therefore a broad unspecific correlation of low $P_T$ modes with a jet. Such a structure has been observed by CMS \cite{CMSBump}, however other mechanism may generate very similar signals.

Given that the precise nature of the source term determines the hydrodynamical response of the medium in an essential way \cite{Solana}, using jet generated shockwaves as a tool to determine the medium speed of sound is therefore a rather ambitious goal which requires substantial further studies to reduce the current uncertainties.

\begin{acknowledgments}
 
This work is supported by the Academy researcher program of the
Academy of Finland, Project No. 130472. 
 
\end{acknowledgments}

\end{document}